\documentclass[12pt,preprint]{emulateapj}
\usepackage{graphicx}

\shorttitle{Transit Exclusion for HD 114762b}
\shortauthors{Stephen R. Kane et al.}
\slugcomment{Submitted for publication in the Astrophysical Journal
  Letters}

\begin{document}

\title{Revised Orbit and Transit Exclusion for HD 114762\lowercase{b}}

\author{
  Stephen R. Kane\altaffilmark{1},
  Gregory W. Henry\altaffilmark{2},
  Diana Dragomir\altaffilmark{1,3},
  Debra A. Fischer\altaffilmark{4},
  Andrew W. Howard\altaffilmark{5,6},
  Xuesong Wang\altaffilmark{7},
  Jason T. Wright\altaffilmark{7,8}
}
\email{skane@ipac.caltech.edu}
\altaffiltext{1}{NASA Exoplanet Science Institute, Caltech, MS 100-22,
  770 South Wilson Avenue, Pasadena, CA 91125}
\altaffiltext{2}{Center of Excellence in Information Systems, Tennessee
  State University, 3500 John A. Merritt Blvd., Box 9501, Nashville,
  TN 37209}
\altaffiltext{3}{Department of Physics \& Astronomy, University of
  British Columbia, Vancouver, BC V6T1Z1, Canada}
\altaffiltext{4}{Department of Astronomy, Yale University, New Haven,
  CT 06511}
\altaffiltext{5}{Department of Astronomy, University of California,
  Berkeley, CA 94720}
\altaffiltext{6}{Space Sciences Laboratory, University of California,
  Berkeley, CA 94720}
\altaffiltext{7}{Department of Astronomy and Astrophysics,
  Pennsylvania State University, 525 Davey Laboratory, University
  Park, PA 16802}
\altaffiltext{8}{Center for Exoplanets \& Habitable Worlds,
  Pennsylvania State University, 525 Davey Laboratory, University
  Park, PA 16802}


\begin{abstract}

Transiting planets around bright stars have allowed the detailed
follow-up and characterization of exoplanets, such as the study of
exoplanetary atmospheres. The Transit Ephemeris Refinement and
Monitoring Survey (TERMS) is refining the orbits of the known
exoplanets to confirm or rule out both transit signatures and the
presence of additional companions. Here we present results for the
companion orbiting HD~114762 in an eccentric 84 day orbit. Radial
velocity analysis performed on 19 years of Lick Observatory data
constrain the uncertainty in the predicted time of mid-transit to
$\sim 5$~hours, which is less than the predicted one-half day transit
duration. We find no evidence of additional companions in this
system. New photometric observations with one of our Automated
Photoelectric Telescopes (APTs) at Fairborn Observatory taken during 
a revised transit time for companion b, along with 23 years of nightly 
automated observations, allow us to rule out on-time central transits 
to a limit of $\sim~0.001$ mag. Early or late central transits are
ruled out to a limit of $\sim~0.002$ mag, and transits with half the 
duration of a central transit are ruled out to a limit of 
$\sim~0.003$ mag.

\end{abstract}

\keywords{planetary systems -- techniques: photometric -- techniques:
  radial velocities -- stars: individual (HD~114762)}


\section{Introduction}
\label{introduction}

The radial velocity (RV) method is still the dominant source of
confirmed exoplanets, although the transit technique is rapidly
producing planet detections. The advantage of transiting planets is
that they allow additional characterization of the planet, namely the
radius and hence the density. For planets transiting bright stars,
atmospheric studies may be undertaken, as was the case for HD~189733b
\citep{knu09a} and HD~149026b \citep{knu09b}. However, most of the
planets detected using transits orbit relatively faint host stars,
making them inaccessible to such investigations. Transit detections
are also biased towards short periods due to the geometric transit
probability \citep{kan08}. The purpose of the Transit Ephemeris
Refinement and Monitoring Survey (TERMS) is to ameliorate these biases
through photometric monitoring of known RV planets at improved times
of predicted transits \citep{kan09}.

The first planet candidate to be detected with the radial velocity
technique orbits the star HD~114762, a late F dwarf. The companion was
discovered by \citet{lat89} who reported the minimum mass as $0.011
\pm 0.001 \ M_\odot$. This discovery was confirmed by \citet{coc91}
and a search for transits by \citet{rob90} ruled out transit depths
greater than 0.01 magnitudes. \citet{hal95} used high-resolution
spectroscopy to measure the projected rotational velocity $v \sin i$
and concluded that the inclination of the companion is likely to be
low, possibly rendering the companion above the limit for deuterium
fusion. There is still uncertainty as to the true inclination of the
companion, particularly with the large range of spin-orbit
misalignments which can occur \citep{fab09}. Further long-term
photometry by \citet{hen97} has found the host star to be
photometrically stable in the $V$ band to 0.001 magnitudes.

Here we present our complete RV dataset from Lick Observatory that has
a time baseline of 19 years. The Keplerian analyses of these data is
used to produce a revised orbital solution that further refines the
period for transit studies and rules out the presence of additional
companions within the system. We also present 23 years of
high-precision photometry from the T2 0.25~m and the T10 0.8~m
Automatic Photoelectric Telescopes (APTs) at Fairborn Observatory. The
most recent data were acquired with the T10 APT whose 1$\sigma$
uncertainty of 0.0012 mag for a single observations is easily
sufficient to find or to rule out transits with a predicted depth of
$\sim 1$\%.


\section{Revised Orbital Parameters}

The RV data were acquired with the 3.0m Shane telescope and the
Hamilton Echelle Spectrograph at Lick Observatory. The data comprise
74 measurements that range from 1990 March through 2009 February and
are shown in Table \ref{rvs}. Of these, 46 were extracted from the
same spectra as used by \citet{but06}. However, the Lick data
reduction pipeline undergoes frequent refinement and so each reduction
is of superior quality to previously published measurements extracted
from those spectra. As data accumulates, the code adaptively
reassesses the relative weights of portions of the spectra based upon
such factors as contamination by weak telluric lines or bad pixels.
The fourth column in Table \ref{rvs} shows the dewar number which was
used with the observation, the relevance being the different CCD
response characteristics which can occur between dewars. This is
accounted for in the Keplerian orbital fitting described below.

\begin{deluxetable}{cccc}
  \tablewidth{0pc}
  \tablecaption{\label{rvs} Lick Radial Velocities}
  \tablehead{
    \colhead{Date} &
    \colhead{Radial Velocity} &
    \colhead{Uncertainty} &
    \colhead{Dewar} \\
    \colhead{(JD -- 2440000)} &
    \colhead{(m\,s$^{-1}$)} &
    \colhead{(m\,s$^{-1}$)} &
    \colhead{}
  }
  \startdata
    7964.963900 &  -185.78 & 22.28 &     6 \\
    8017.833900 &   315.12 & 20.41 &     6 \\
    8018.840200 &   277.92 & 17.40 &     6 \\
    8019.830500 &   289.41 & 27.20 &     6 \\
    8375.852800 &  -600.95 & 14.63 &     6 \\
    8437.738000 &   262.91 & 19.82 &     6 \\
    8649.091100 &   465.42 & 17.25 &     8 \\
    8670.973032 &   569.49 & 21.00 &     8 \\
    9068.820700 &   452.17 & 25.78 &     8 \\
    9068.842500 &   505.42 & 30.66 &     8 \\
    9096.814400 &   518.85 & 27.31 &     8 \\
    9096.835900 &   532.71 & 28.77 &     8 \\
    9114.776700 &    76.31 & 21.17 &     8 \\
    9174.705300 &   573.39 & 14.70 &     8 \\
    9350.008000 &   476.46 & 17.36 &     8 \\
    9375.011500 &  -296.95 & 17.90 &     8 \\
    9464.786200 &  -636.28 & 25.35 &     8 \\
    9469.781200 &  -554.16 & 16.03 &     8 \\
    9768.932800 &   496.52 & 25.17 &    39 \\
    9768.955078 &   513.54 & 22.34 &    39 \\
    9801.853516 &  -626.62 & 24.68 &    39 \\
    9801.875000 &  -602.62 & 28.81 &    39 \\
    9801.899414 &  -597.07 & 24.61 &    39 \\
    9802.825195 &  -535.91 & 23.30 &    39 \\
    9802.848633 &  -556.53 & 20.99 &    39 \\
    9802.871094 &  -585.50 & 23.00 &    39 \\
    9803.818359 &  -572.41 & 21.69 &    39 \\
    9803.840820 &  -600.02 & 22.14 &    39 \\
    9803.863281 &  -578.70 & 24.69 &    39 \\
    9858.733398 &   409.48 & 25.35 &    39 \\
    9881.718750 &  -445.90 &  9.60 &    39 \\
    9892.701172 &  -304.06 & 24.17 &    39 \\
    9892.723633 &  -317.33 & 26.76 &    39 \\
    9892.746094 &  -336.66 & 27.65 &    39 \\
    9893.698242 &  -218.52 & 27.55 &    39 \\
    9893.721680 &  -271.28 & 24.85 &    39 \\
    9893.744141 &  -256.03 & 25.36 &    39 \\
    9894.697266 &  -237.50 & 20.05 &    39 \\
    9894.719727 &  -170.63 & 18.06 &    39 \\
    9894.742188 &  -203.70 & 19.28 &    39 \\
    9913.755859 &   585.09 &  9.80 &    39 \\
    9914.675781 &   608.46 &  9.31 &    39 \\
   10072.061523 &   355.86 & 17.98 &    39 \\
   10120.954102 &   151.07 & 23.88 &    39 \\
   10120.976562 &   132.44 & 25.13 &    39 \\
   10121.968750 &    96.44 & 23.06 &    39 \\
   10121.990234 &    92.78 & 24.77 &    39 \\
   10124.940430 &    -7.39 & 23.47 &    39 \\
   10124.962891 &     1.09 & 26.03 &    39 \\
   10126.963867 &  -144.75 & 26.88 &    39 \\
   10126.984375 &   -96.89 & 39.75 &    39 \\
   10128.038086 &  -257.56 & 24.87 &    39 \\
   10128.062500 &  -167.79 & 26.50 &    39 \\
   10128.970703 &  -245.66 & 22.94 &    39 \\
   10128.994141 &  -212.46 & 20.20 &    39 \\
   10144.940430 &  -335.55 & 11.58 &    39 \\
   10172.857422 &   716.70 & 51.27 &    39 \\
   10181.819336 &   583.34 & 14.87 &    39 \\
   10187.833008 &   566.28 & 23.11 &    39 \\
   10187.856445 &   520.18 & 26.32 &    39 \\
   10200.783203 &   289.36 & 23.24 &    39 \\
   10200.805664 &   284.61 & 22.44 &    39 \\
   10504.960938 &   623.54 & 17.04 &    39 \\
   11628.850586 &   197.24 & 13.89 &    18 \\
   12033.864258 &   521.20 & 10.93 &    18 \\
   12833.715820 &   -71.10 &  9.78 &    24 \\
   13068.975586 &  -392.18 & 14.66 &    24 \\
   13132.822266 &   373.24 & 10.72 &    24 \\
   13389.043333 &   270.11 & 12.56 &    24 \\
   13544.723507 &   501.86 & 10.65 &    24 \\
   13545.745370 &   494.24 & 10.29 &    24 \\
   14165.945718 &  -625.59 & 11.30 &    24 \\
   14547.955891 &   527.27 & 10.38 &    24 \\
   14865.043218 &   576.89 & 13.83 &    24
   \enddata
\end{deluxetable}

We used Spectroscopy Made Easy \citep{val96} to fit high-resolution
Lick spectra of HD~114862, applying the wavelength intervals, line
data, and methodology of \citet{val05}. We further constrained the
surface gravity using Yonsei-Yale (Y$^2$) stellar structure models
\citep{dem04} and revised \textit{Hipparcos} parallaxes \citep{van07} 
with the iterative method of \citet{val09}. The resulting stellar
parameters listed in Table \ref{stellar} are effective temperature,
surface gravity, iron abundance, projected rotational velocity, mass,
and radius. The stellar radius, $R_\star = 1.24 R_\odot$, is crucial
for estimating the depth and duration of a planetary transit. These
properties are consistent with a very metal-poor early-G sub-giant.

\begin{deluxetable}{lc}
  \tablecaption{\label{stellar} Stellar Properties}
  \tablehead{
    \colhead{Parameter} &
    \colhead{Value}
  }
  \startdata
  $V$                       & 7.3 \\
  $B-V$                     & 0.525 \\
  Distance (pc)             & $38.7 \pm 1.1$ \\
  $T_\mathrm{eff}$ (K)      & $5673 \pm 44$ \\
  $\log g$                  & $4.135 \pm 0.060$ \\
  $[$Fe$/$H$]$              & $-0.774 \pm 0.030$ \\
  $v \sin i$ (km\,s$^{-1}$) & $1.77 \pm 0.50$ \\
  $M_\star$ ($M_\odot$)     & $0.83 \pm 0.01$ \\
  $R_\star$ ($R_\odot$)     & $1.24 \pm 0.05$
  \enddata
\end{deluxetable}

We fit a single-planet Keplerian orbital solution to the RV data using
the techniques described in \citet{how10} and the partially
linearized, least-squares fitting procedure described in
\citet{wri09}. This was performed both with and without the inclusion
of a linear trend in order to determine the significance of including
that free parameter. The parameter uncertainties are extracted from
the sampling distribution of each parameter through a non-parametric
bootstrap analysis \citep{fre81}. Table \ref{planet} lists the fit
parameters both with and without the trend.

\begin{deluxetable}{lcc}
  \tablecaption{\label{planet} Keplerian Fit Parameters}
  \tablehead{
    \colhead{Parameter} &
    \colhead{Trend} &
    \colhead{No Trend}
  }
  \startdata
  $P$ (days)                       & $83.9152 \pm 0.0028$  & $83.9151 \pm 0.0030$ \\
  $T_c\,^{a}$ (JD -- 2440000)      & $15326.595 \pm 0.242$ & $15326.665 \pm 0.242$ \\
  $T_p\,^{b}$ (JD -- 2440000)      & $9889.141 \pm 0.190$  & $9889.106 \pm 0.186$ \\
  $e$                              & $0.3325 \pm 0.0048$   & $0.3354 \pm 0.0048$ \\
  $K$ (m\,s$^{-1}$)                & $612.72 \pm 3.35$     & $612.48 \pm 3.52$ \\
  $\omega$ (deg)                   & $201.41 \pm 1.04$     & $201.28 \pm 1.01$ \\
  $dv/dt$ (m\,s$^{-1}$\,yr$^{-1}$) & $-3.83 \pm 1.91$      & N/A \\
  $M_p \sin i$ ($M_J$)             & $10.99 \pm 0.09$      & $10.98 \pm 0.09$ \\
  $a$ (AU)                         & $0.353 \pm 0.001$     & $0.353 \pm 0.001$ \\
  $\chi^2_{\mathrm{red}}$          & 1.29                  & 1.32 \\
  rms (m\,s$^{-1}$)                & 27.53                 & 27.42
  \enddata
  \tablenotetext{a}{Time of mid-transit.}
  \tablenotetext{b}{Time of periastron passage.}
\end{deluxetable}

%

\begin{figure}
  \includegraphics[width=8.5cm]{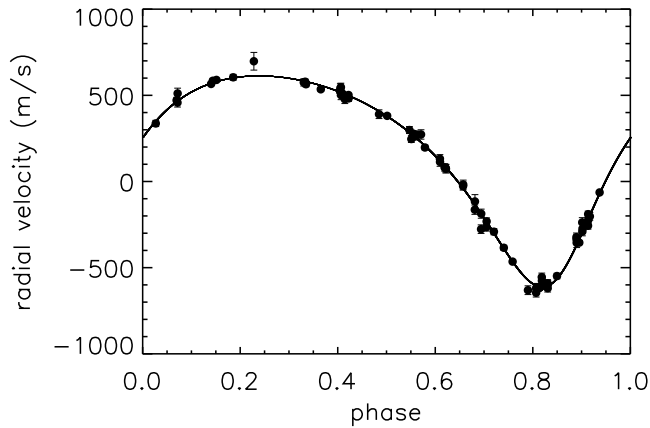}
  \caption{Radial velocity data of HD~114762 acquired with the 3.0m
    Shane telescope at Lick Observatory. Also shown is the best-fit
    Keplerian model for the planet.}
  \label{rv}
\end{figure}

As indicated by the $\chi^2_{\mathrm{red}}$ and the rms scatter of the
residuals, the Keplerian fit which includes the trend does not improve
the fit given the addition of the free parameter. Thus this trend is
unlikely to be due to any additional companions within the system.
This is consistent with the findings of \citet{mug06} who exclude
companions with masses greater than 66~$M_J$ at orbital radii of
316--2674~AU. Note that \citet{pat02} detect a stellar companion to
HD~114762, further characterized by \citet{bow09}, at an angular
separation of 3.3\arcsec (180~AU). This companion could feasibly cause
the observed trend, but the significance level is too low for us to
claim such a detection. We therefore adopt the solution without the
trend included, shown in Figure \ref{rv}. The fit required four
additional free parameters due to the offsets induced by the changing
of the detector dewars. The offsets with respect to the observations
acquired with dewar 6 are $-6.81 \pm 5.88$, $19.31 \pm 6.26$, $17.42
\pm 8.91$, and $-5.40 \pm 7.12$ m\,s$^{-1}$ for dewars 8, 39, 18, and
24 respectively. The uncertainty in the period for the presented
orbital solution is a factor of 4 improvement over the previous
solution by \citet{but06} and allows a more robust transit search to
be carried out.



\section{Transit Ephemeris Refinement}

From the stellar and planetary properties listed in Tables
\ref{stellar} and \ref{planet}, we derive a planetary radius of $R_p =
1.11 R_J$ using the methods described in \citet{bod03}. This results
in a predicted transit duration of 0.49 days and a predicted transit
depth of 0.91\%. The uncertainty in the stellar mass/radius and
subsequent uncertainty in the planetary mass/radius have a minor
effect on the estimated transit duration but in no way affects the
predicted transit mid-points since these are derived from the orbital
parameters \citep{kan09}. Based upon the revised orbital parameters,
we computed a new transit ephemeris, applied in the following
section.

As described by \citet{kan08}, the probability of a planetary transit
is a strong function of both the eccentricity and the argument of
periastron and is at a maximum when the periastron passage occurs
close to the star-observer line of sight ($\omega = 90\degr$). The
measured periastron argument for HD~114762b leads to an orbital
configuration which favors a secondary eclipse rather than a primary
transit. The transit probability based upon the parameters of Table
\ref{planet} is 1.71\%. For comparison, an equivalent circlular orbit
would have a transit probability of 1.73\%, but the same eccentricity
with a periastron argument of $90\degr$ would have a transit
probability of 2.60\%.


\section{Transit Exclusion and Implications}

As part of our long-term program to measure luminosity cycles in
solar-type stars \citep{hen99}, we have acquired 23 consecutive years
of photometric observations of HD~114762 with our T2 0.25~m and T10
0.80~m automatic photometric telescopes (APTs) at Fairborn Observatory
in southern Arizona. The T2 APT used a photodiode detector to make
differential measurements of stars through Johnson $VRI$ filters. It
collected 1989 nightly observations of HD~114762 during the 1989
through 2001 observing seasons. Details concerning the data
acquisition and reduction with this telescope can be found in
\citet{hen95} and \citet{per00}. The T10 APT has a two-channel
photometer with a dichroic filter that separates the Stro\"mgren $b$
and $y$ photometric bands so they can be measured simultaneously by
two EMI 9124QB photomultiplier tubes. T10 collected 1137 nightly
observations of HD~114762 during the 2002 through 2011 observing
seasons. We have averaged the Stro\"mgren $b$ and $y$ band
measurements from T10 to create a new $(b+y)/2$ ``passband'' that
gives us improved precision. T10 operation is very similar to our T8
0.80~m APT, described in \citet{hen95,hen99}.

\begin{figure*}
  \begin{center}
    \begin{tabular}{cc}
      \includegraphics[width=8.2cm]{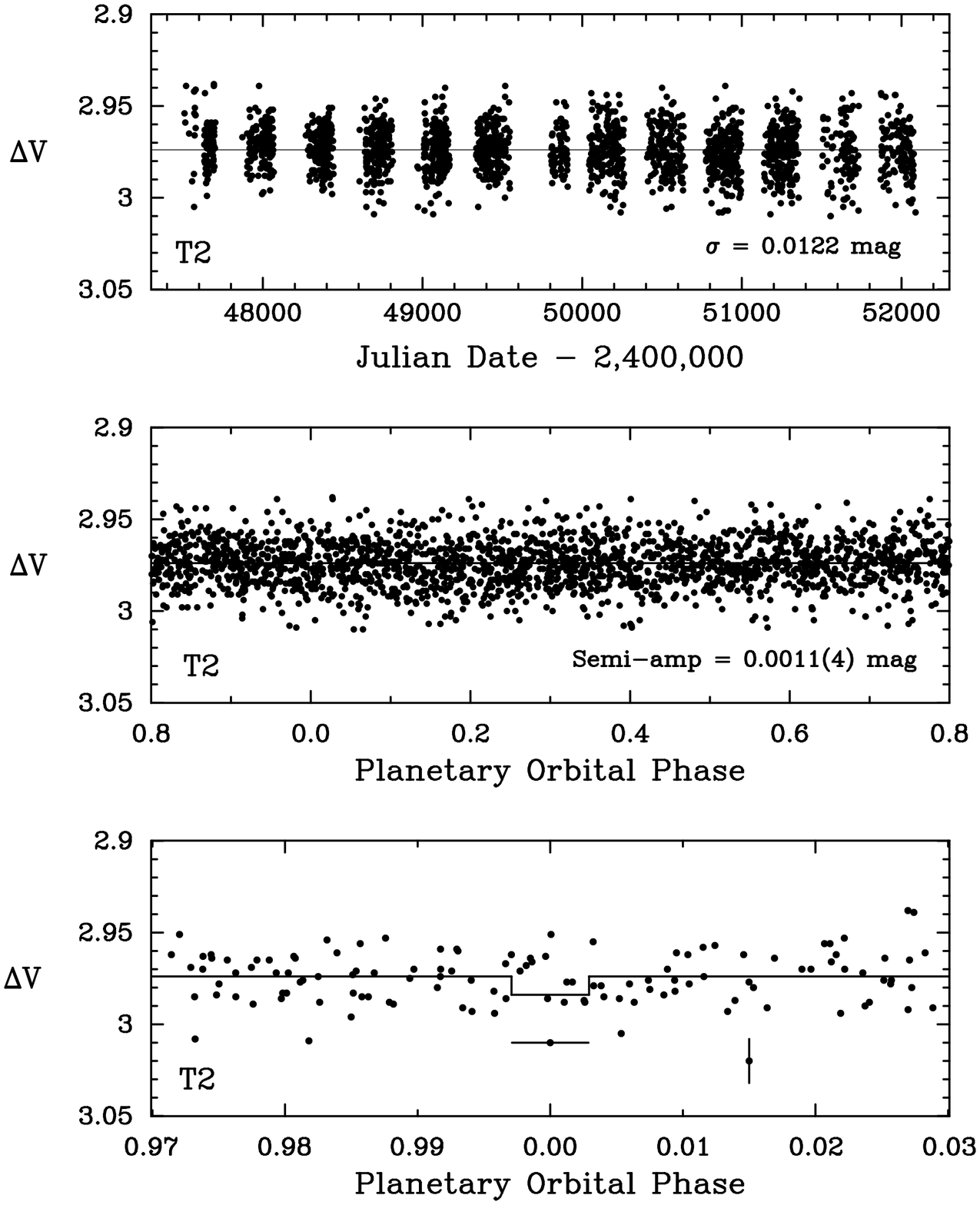} &
      \includegraphics[width=8.2cm]{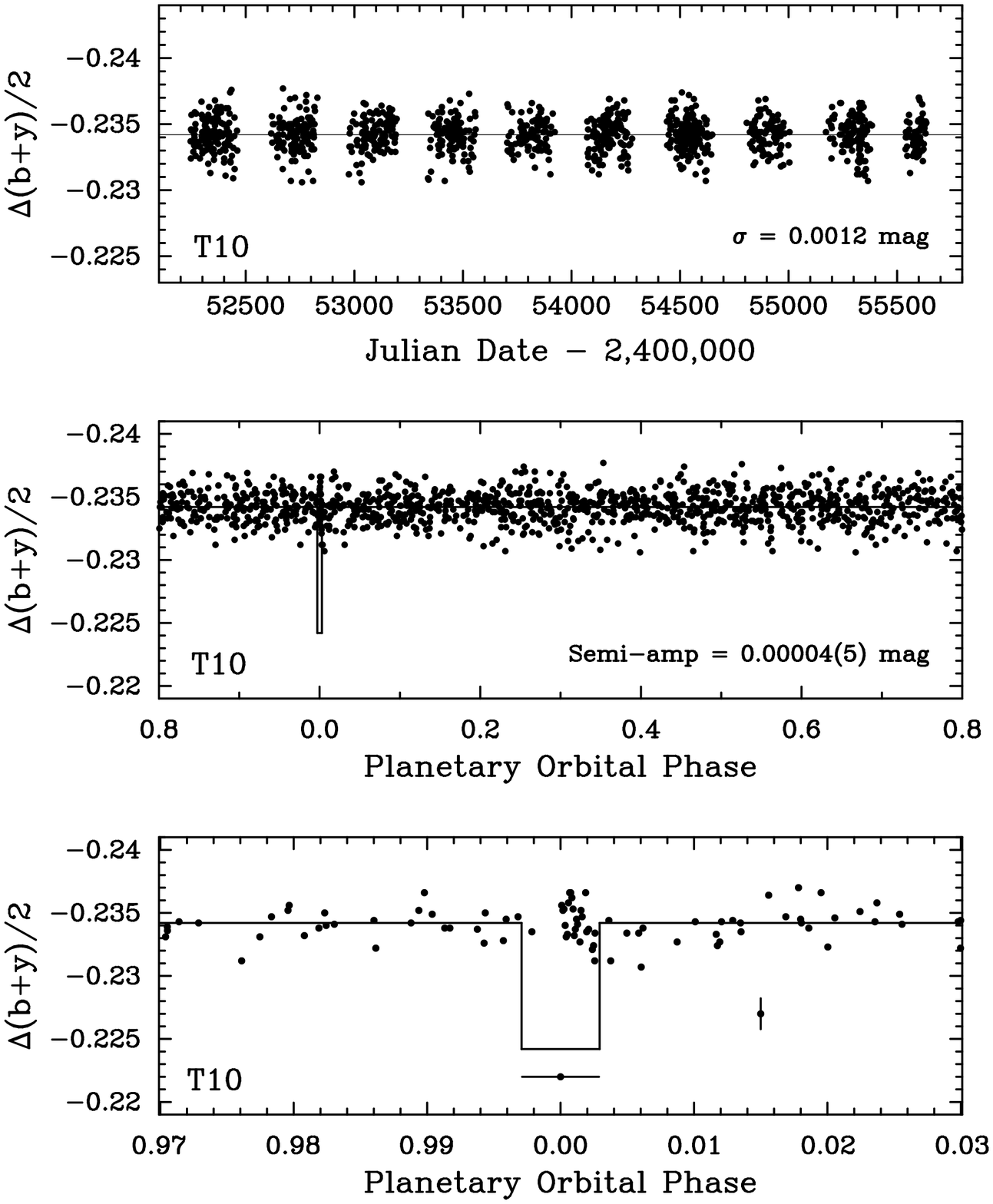}
    \end{tabular}
  \end{center}
  \caption{Top panels: The 1989 T2 APT $\Delta V$ observations (left)
    and the 1137 T10 APT $\Delta (b+y)/2$ observations (right) of
    HD~114762 collected during 23 consecutive years between 1989 and
    2011. Middle panels: The T2 and T10 observations plotted modulo
    the revised RV period, $P$ (Table \ref{planet}) with phase 0.0
    equal to the predicted time of mid-transit, $T_c$ (Table
    \ref{planet}). The observations place very tight limits on
    photometric variability on the radial velocity period. Bottom
    panels: The measurements near phase 0.0 plotted on an expanded
    scale. The solid curve shows the predicted depth ($\sim~1\%$) and
    duration (0.49 days) of a central transit computed with our
    revised orbital parameters (Table \ref{planet}).  The horizontal
    error bar under the transit window gives the $\pm~1\sigma$
    uncertainty in $T_c$. Our results show that on-time, central
    transits with a depth of 0.001 mag or more do not occur.}
  \label{photometry}
\end{figure*}

The T2 $\Delta V$ and T10 $\Delta (b+y)/2$ observations are shown in
the left and right panels, respectively, of Figure
\ref{photometry}. Both data sets have been normalized so that all
seasonal means agree with the mean of the first year, shown as the
horizontal line in the two top panels. This removes any long-term
brightness variability in the comparison stars and in HD~114762
itself; corrections were typically a few tenths of a mmag. The top
panels of Figure \ref{photometry} give for each APT the standard
deviation of a single observation from the mean of the whole data set:
0.0122 mag and 0.0012 mag for T2 and T10, respectively. Both values
are consistent with the precision of a single observation from each
telescope, establishing the order of magnitude better precision of T10
over T2. Thus, we find that HD~114762 is constant to high precision
on both night-to-night and year-to-year timescales.

The middle panels of Figure~\ref{photometry} plot the two data sets
with respect to the orbital phase of the companion, with phase 0.0
equal to the predicted time of mid-transit, $T_c$, calculated from 
our revised orbital parameters in Table \ref{planet}. A least-squares 
sine fit to each data set gives semi-amplitudes of $0.0011~\pm~0.0004$ 
and $0.00004~\pm~0.00005$ mag. This confirms to very high precision the
lack of stellar activity that might otherwise mimic periodic radial
velocity variations in HD~114762 (see, e.g., \citet{que01,pau04,bon07,
for09}).

The bottom panels show the portion of the middle panels around phase
0.0 plotted with an expanded scale on the x-axis.  The solid curve in
each panel shows the predicted time ($T_c$), depth ($\sim~1\%$), and
duration (0.49 days) of a central transit, computed from the new
orbital and planetary parameters derived above. The horizontal error
bar below the transit shows the $\pm~1\sigma$ uncertainty in
$T_c$. With the present radial velocity data set in this paper, the
uncertainty in $T_c$ is almost exactly equal to the transit
duration. For the T2 observations in the bottom left panel of
Figure~\ref{photometry}, the mean of the 12 in-transit observations
agrees with the mean of the 1977 out-of-transit observations to
$0.0001~\pm~0.0035$ mag, where the uncertainty is calculated by adding
in quadrature the uncertainty of each of the two means. The
uncertainty of the in-transit mean level dominates the total
uncertainty because of the relatively fewer observations in transit
than out.  We have thus measured the difference between the in-transit
and out-of-transit light levels to a precision of 0.0035 mag. If we
assume the need for a $3\sigma$ event to ensure detection, the T2
observations can only rule out transits 0.0105 mag ($\sim~1\%$) or
deeper. Therefore, the T2 observations have just enough sensitivity to
detect the predicted transits but fail to do so.

For the much more precise T10 observations in the bottom right panel
of Figure~\ref{photometry}, the difference between the mean of the 26
in-transit observations and the mean of the 1111 out-of-transit
observations is $0.0001~\pm~0.0003$ mag. Therefore, transits with a
$3~\sigma$ depth of 0.0009 mag or more should be detectable. We round
this to 0.001 mag and note that this limit applies only to
full-duration, on-time events. We must make this distinction because
of the non-random distribution of observations in the predicted
transit window. We have only one observation very early in the first
half of the transit window and 25 observations that span the second
half. Twenty-two of these 25 points came from a single night
(JD~2455326) when we monitored HD~114762 at a higher cadence. The
worst case for the detection of full-duration transits occurs if they
are 1 standard deviation early, i.e., the window slides leftward
one-half of its width. In this case, we loose all of the monitoring
observations from the window and pick up only five nightly
observations. With only $\onequarter$ the number of observations in
the transit window, the $3~\sigma$ transit depth limit doubles to
0.002 mag. Thus, we can confidently rule out early or late
full-duration transits to a limit of 0.002 mag. If the companion does
transit as described, this is equivalent to ruling out radii of $0.54
R_J$, or a density of 86~g\,cm$^{-3}$. For comparison, the density of
the planet HAT-P-20b, with a mass of $7.246 M_J$ and a radius of
$0.867 R_J$, is 13.8 ~g\,cm$^{-3}$.

It is also worth considering transits whose trajectory lies along a
chord of the star approximately half the stellar diameter. An on-time
transit along such a chord would contain half the number of
observations, all during the second half of the shorter transit.
Thus, the detectable limit of an on-time transit with $\onehalf$ the
full duration increases from 0.001 mag to $\sim~0014$ mag. The worst
case for detection of these shorter transits occurs if they are
$\onehalf~\sigma$ early. In this case, the transit window contains
only one observation, namely, the first observation in the transit
window in Figure~\ref{photometry}. The standard deviation of a single
observation is 0.0012 mag, so the $3~\sigma$ depth limit for this
worst case is $\sim~0.0036$ mag. But this worst case is very unlikely
and happens only for $T_c$ exactly $\onehalf~\sigma$ early. A little
earlier or later, and the transit window begins picking up additional
observations, and the depth limit becomes tighter. Therefore, we can
exclude transits with $\onehalf$ full duration to a limit of
$\sim~0.003$ mag. Similarly to that described above, this rules out
transits of companions with densities lower than 47~g\,cm$^{-3}$.

The T10 observations in the lower-right panel illustrate the utility
of the TERMS approach to finding transits of planets with intermediate
orbital periods. Of the 26 in-transit observations, only 4 of the
nightly observations taken over 13 observing seasons fell at random
within the transit window. The other 22 in-transit observations were
acquired in a single night, 2010 May 10 UT, predicted from our updated
orbit based on our new radial velocities acquired within the TERMS
program.


\section{Conclusions}

These new RV data refine the orbit of the first planetary candidate
discovered to orbit a solar-type star. Our analysis of Lick spectra
have greatly improved the estimated parameters of the host
star. Including a linear trend into the Keplerian orbital fit does not
significantly improve the rms scatter of the residuals and we thus
find no evidence of any additional companions in the system. From the
Keplerian fit we produce a refined transit ephemeris onto which we
have folded 23 years of precise photometry acquired with the APT at
Fairborn Observatory. Our photometric observations rule out on-time
central transits to a limit of $\sim~0.001$ mag, early or late central
transits to a limit of $\sim~0.002$ mag, and transits of $\onehalf$
full duration to a limit of $\sim~0.003$ mag. Transits shorter than
$\onehalf$ full duration are possible to hide in the gap in our
transit coverage but are highly unlikely due to the low probability of
the required inclination angles and the exact timing needed to fit the
transit within the gap. The absence of planetary transits constrains
the orbital inclination to be less than $89.0\degr$, which raises the
minimum mass by a neglible amount. For the companion to transit
undetected for these data, the density would need to be in excess of
47~g\,cm$^{-3}$, which is well outside the expected range for the
planetary and brown dwarf regimes. The absolute confirmation of the
planetary nature of the companion will likely need to await precision
astrometric observations of this target to determine the orbital
inclination.


\section*{Acknowledgements}

The authors would like to thank David Latham for useful comments on
the manuscript. We would also like to thank the anonymous referee,
whose comments greatly improved the quality of the paper. The Center
for Exoplanets and Habitable Worlds is supported by the Pennsylvania
State University, the Eberly College of Science, and the Pennsylvania
Space Grant Consortium.  G.W.H. acknowledges long-term support from
NASA, NSF, Tennessee State University, and the State of Tennessee
through its Centers of Excellence program.


\end{document}